\documentclass[fleqn,usenatbib]{mnras}

\usepackage{newtxtext,newtxmath}
\usepackage[T1]{fontenc}

\DeclareRobustCommand{\VAN}[3]{#2}
\let\VANthebibliography\thebibliography
\def\thebibliography{\DeclareRobustCommand{\VAN}[3]{##3}\VANthebibliography}

\usepackage{graphicx}	% Including figure files
\usepackage{amsmath}	% Advanced maths commands
 \usepackage{multirow}
\title[Monitoring NIR variability in NGC~4388]{A Decade of Near-Infrared Variability in NGC4388: Insights into the AGN Structure}

\author[L. G. Dahmer-Hahn et al.]{
Luis G. Dahmer-Hahn,$^{1}$\thanks{E-mail: luisgdh@shao.ac.cn}, Alberto Rodr\'iguez-Ardila$^{2,3}$\thanks{aardila@lna.br }, Marina Bianchin$^{4, 5}$, Rogemar A. Riffel$^{4}$,   \newauthor Rog\'erio Riffel$^{6, 7}$, Thaisa Storchi-Bergmann$^{6}$, Lei Hao$^{1}$
\\
$^{1}$Key Laboratory for Research in Galaxies and Cosmology, Shanghai Astronomical Observatory, Chinese Academy of Sciences,  80 Nandan Road, \\ Shanghai 200030, China\\
$^{2}$Laborat\'orio Nacional de Astrof\'isica. Rua dos Estados Unidos, 154, 37504-364 Itajub\'a, MG, Brazil\\
$^{3}$Instituto Nacional de Pesquisas Espaciais, Av. dos Astronautas, 1758 - Jardim da Granja S\~ao Jos\'e dos Campos/SP - 12227-010, Brazil.\\
$^{4}$Departamento de F\'isica, Centro de Ci\^encias Naturais e Exatas, Universidade Federal de Santa Maria, 97105-900, Santa Maria, RS, Brazil\\
$^{5}$Department of Physics and Astronomy, 4129 Frederick Reines Hall, University of California, Irvine, CA 92697, USA\\
$^{6}$Instituto de F\'isica, Universidade Federal do Rio Grande do Sul, Av. Bento Gon\c{c}alves 9500, 91501-970 Porto Alegre, RS, Brazil\\
$^{7}$Instituto de Astrof\'isica de Canarias, E-38200 La Laguna, Tenerife, Spain
}

\date{Accepted 2023 June 9. Received 2023 June 8; in original form 2023 April 30}

\pubyear{2022}

\begin{document}
\label{firstpage}
\pagerange{\pageref{firstpage}--\pageref{lastpage}}
\maketitle

\begin{abstract}

Variability studies have proven to be a powerful diagnostic tool for understanding the physics and properties of of Active Galactic Nuclei (AGNs). They provide insights into the spatial and temporal distribution of the emitting regions, the structure and dynamics of the accretion disk, and the properties of the central black hole. Here, we have analysed the K-band spectral variability of the Seyfert~1.9/2 galaxy NGC~4388 spanning five epochs over a period of ten years. We have performed spectral synthesis of the nuclear region and found that the contribution of warm dust (T$\sim$800~K) declined by 88~\% during these 10 years. In the same period, the [\ion{Ca}{VIII}] coronal line decreased 61~\%, whereas Br$\gamma$ emission declined 35$\%$. For the \ion{He}{I} and H$_2$, we did not detect any significant variation beyond their uncertainties. Based on the time span of these changes, we estimate that the region where the warm dust is produced is smaller than 0.6~pc, which suggests that this spectral feature comes from the innermost part of the region sampled, directly from the AGN torus. On the other hand, the bulk of [\ion{Ca}{VIII}] is produced in the inner $\sim$2~pc and the nuclear Br$\gamma$ region is more extended, spanning a region larger than 3~pc. Lastly, \ion{He}{I} and H$_2$ are even more external, with most of the emission probably being produced in the host galaxy rather than in the AGN. This is the first spectroscopic variability study in the NIR for an AGN where the central source is not directly visible.
\end{abstract}


%%%%%%%%%%%%%%%%%%%%%%%%%%%%%%%%%%%%%%%%%%%%%%%%%%

%%%%%%%%%%%%%%%%% BODY OF PAPER %%%%%%%%%%%%%%%%%%

\section{Introduction}

Active Galactic Nuclei (AGNs) are among the most energetic and luminous objects in the universe, reaching integrated luminosities of up to 10$^{48}$~erg$\cdot$s$^{-1}$ \citep{Koratkar&Blaes99,Bischetti+17}. They are powered by the accretion of matter onto supermassive black holes at their centres, which can release huge amounts of energy across the electromagnetic spectrum \citep[e.g.][]{Netzer15,Padovani+17,StorchiBergmann&SchnorrMuller19}. AGNs display a wide range of observational properties, including broad emission lines, continuum radiation spanning a broad range of wavelengths, and often exhibit variability on different timescales \citep{Gaskell+03,Peterson+01, Padovani+17}.

AGNs have been found to be variable at all wavelengths at which they have been observed \citep{Peterson+01}, from $\gamma$-rays \citep[e.g.][]{Gaidos+96} and X-rays \citep[e.g.][]{McHardy01, Sanfrutos+16, Mehdipour+17}, going through visible light \citep{Winge+95,Winge+96,Burke+21}, up to longer wavelengths such as infrared \citep[e.g.][]{Koshida+14,Sanchez+17} and radio \citep[e.g.][]{Hovatta+07}. Depending on the properties of the AGN, this variability can happen in very different timescales, ranging from minutes to decades, with different timescales and wavelength ranges being attributed to different physical processes.

Variability studies in AGNs have been essential for advancing our understanding of the physical processes that occur in these objects. For example, studies of AGN variability have been used to probe the structure and dynamics of the accretion disk. They have also been used to investigate the mass and spin of the central black hole \citep[e.g.][]{McHardy+06,Cackett+13,McHardy+14,Emmanoulopoulos+14}.

It is expected that the variability of the central source will leave fingerprints on high ionisation emission lines, such as the coronal lines, as well as on the hot dust emission from the clumpy torus, since they are directly associated with the AGN emission. Despite offering insight into so many interesting features, variability studies on the near-infrared (NIR) are rare. \citet{Koshida+14} presented a dust reverberation survey for 17 nearby Seyfert~1 galaxies. They found a delayed response of the K-band light curve after the V-band light curve ranging between 9 and 170~days for all targets. They also found that these lag times strongly correlate with their optical luminosity. However, photometric surveys such as these are technically limited, since they cannot distinguish between emission lines and continuum emission. This distinction gives us insights into the geometry of the central source, as well as into parameters such as flux and dust temperature.

%\citet{Sanchez+17} obtained photometric NIR light curves of over 2,000 sources in an 1.5~deg$^2$ area. Through their 5-year period, they were able to find that AGNs exhibiting broad lines (BL) have a considerably larger fraction of variable sources than targets with only narrow lines (NL). They also reported that most of the low luminosity variable NL sources correspond to BL AGN, where the host galaxy could dampen the variability signal. For high luminosity variable NL, on the other hand, they suggested that there can be examples of true type II AGN, or BL AGN with limited spectral coverage which results in missing the BL emission.

In their study, \citet{Sanchez+17} analysed the photometric near-infrared (NIR) light curves of more than 2,000 sources within a 1.5~deg$^2$ area. Over a period of five years, they discovered that active galactic nuclei (AGNs) that exhibit broad lines (BL) have a significantly higher proportion of variable sources compared to those with only narrow lines (NL). The authors also found that most of the low-luminosity variable NL sources are likely BL AGN, where the variability signal may be weakened by the host galaxy. In contrast, for high-luminosity variable NL, the authors suggested that these could be examples of true type II AGN or BL AGN with limited spectral coverage, which may result in the BL emission being missed.

So far, NIR spectroscopic reverberation mapping campaigns have been conducted in only two Seyfert~1 galaxies, NGC\,5548 \citep{Landt+19} and Mrk~876 \citep{Landt+23}. For NGC~5548, the authors reported that both the accretion disc and the hot dust contribute to the variability, with lag times of $\sim$70 days. Also, for Mrk~876, they found that the luminosity-based dust radii are larger than the dust response time obtained by a contemporaneous photometric reverberation mapping campaign, by a factor of $\sim$2. This result is well explained by a flared, disk-like structure for the hot dust. In both cases, the authors detected dust emission components with temperatures ranging from 1000 to 1500~K.

Such variability studies in AGNs are powerful diagnostic tools for understanding the physics and properties of these objects, providing insights into the spatial and temporal distribution of the emitting regions, the structure and dynamics of the accretion disk, and the properties of the central black hole. Thus, variable AGNs are key to the understanding of geometrical properties of the inner region.

Generally speaking, the more luminous the AGN, the more dramatic its variability can be \citep{Peterson+01}. The main reason for this is that in more luminous sources we have access to smaller and more internal regions, which can vary more easily than bigger and more external areas. Also, it is easier to identify the variability in highly luminous sources. This makes variable, but not-so-bright AGNs (such as Seyfert~2) rare but very important when probing the physics of the circumnuclear regions. 

%\citet{Hernandez-Garcia+15} monitored 26 Seyfert~2 using X-ray data, in timescales ranging from a few hours to years. They found that, for 25 objects with long-term data, 11 show long-term variation. On the other hand, short-term variability was probed in ten cases, but variations were not found. They reported that these variations are mainly driven by changes in the nuclear power (nine cases), while variations at soft energies or related to absorbers at hard X-rays are less common.

\subsection{NGC 4388}

NGC 4388 is an SA(s)b galaxy in the constellation of Virgo \citep{Veilleux+99}, at a distance of 19~Mpc \citep{Kuo+11}. It is one of the brightest galaxies of the Virgo Cluster due to its Seyfert\,1.9/2 nucleus \citep{Forster+99}. Also, it is nearly edge-on, with an inclination of 79$^\circ$ \citep{Damas-Segovia+16}.

The interstellar medium of the galaxy has undergone a ram-pressure stripping event $\sim$200\,Myr ago, leading to a recent quenching of the star formation activity in the outer, gas-free galactic disk \citep{Damas-Segovia+16,Vollmer+18}. Due to its highly eccentric orbit within the Virgo cluster, the galaxy passed close to the cluster centre, which led to the loss of much of its neutral hydrogen caused by the interaction with the inter-cluster medium.

Because of its water maser emission from a circumnuclear disc, the mass of its supermassive black hole (SMBH) can be accurately measured. \citet{Kuo+11} studied the kinematics of this water maser by employing very long baseline interferometry, also deriving a SMBH mass of 8.5$\pm$0.2$\times$10$^6M_\odot$.

Ionised gas extending up to 35~kpc has been found around the galaxy \citep{Yoshida+02}. The same authors reported that the [\ion{O}{III}]/H$\alpha$ map indicates that the inner 12\,kpc may be excited by  AGN radiation. Also, although the excitation mechanism of the outer region is unclear, it is likely that the nuclear radiation is also a dominant source of its ionisation.

This galaxy also has a double-peaked radio jet \citep{Stone+88,Hummel&Saikia91}, with a primary peak on the nucleus and a secondary peak 230 pc southwest of it. The same authors also reported the detection of a plume of radio plasma to the north of its nucleus.

In the NIR, this target exhibits complex spatially-extended emission lines. According to \citet{RodriguezArdila+17}, who analysed this object through NIR spectroscopy, radiation from the central source alone cannot explain the observed high-ionisation lines. The additional excitation most likely comes from shocks between the radio jet and the ambient gas. 

This source also shows X-ray variability properties. \citet{Fedorova+11} monitored this source between 2003 and 2009 using INTEGRAL and Swift data, and found slow strong variations of the hard X-ray emission. According to their analysis, the flux variability on timescales of 3–6 months in the 20-60 keV energy range is of the order of $\sim$2. They also detected significant spectral shape changes of the 20–300 keV spectrum, uncorrelated with the flux level. Lastly, they found that the lower value of the exponential cut-off at high energies is similar to that of radio-loud AGN, even though NGC~4388 is a radio-quiet object.

All of the above makes this galaxy an ideal target to study AGN properties. As a galaxy with confirmed variability, we are able to assess the geometrical properties of its AGN. Also, since its central source is not directly accessible in the NIR, we are able to probe its circumnuclear properties without the broad line region overshadowing the more extended regions. 

%We show in upper left panel of Fig~\ref{fig:NGC4388}, an optical image\footnote{Publicly available at https://esahubble.org/images/potw1649a/} of the galaxy obtained with the Hubble Wide Field Camera 3 (WFC3). In white, we show the orientation, as well as the angular and spatial scale for this object. In black, we show the zoom in region corresponding to the bottom panels. On the bottom panels, we show in black radio contours at 5\,cm \citep{Damas-Segovia+16}, overlaid on Hubble (left) and H$\alpha$ \citep{Rossa&Dettmar03} (right) images. In red, we show the field of view (FoV) of our NIR datacubes. In the upper right panels, we show from left to right our J+K NIFS continuum, log of H$\alpha$, and in the last panel the log of [\ion{O}{III}] \citep{Schmitt+03}.

In this paper, we investigate the variability of this target in the NIR by employing spectroscopic data obtained spanning a period of 10 years. This paper is structured as follows: in section~\ref{sect:data} we discuss observation and reduction of our data, in section~\ref{sect:cont} we discuss the variability at the continuum level, in section~\ref{sect:elines} we discuss the variability of the emission lines, and in section~\ref{sect:remarks} we summarise our results and give our final remarks.

% \begin{figure}
%     \centering
%     \includegraphics[width=\columnwidth]{NGC4388_1.pdf}
%     \caption{Hubble Wide Field Camera 3 (WFC3) optical image of NGC\,4388, publicly available at \url{https://esahubble.org/images/potw1649a/}. In white we show the orientation of the image, and the angular scale of the galaxy.}
%     \label{fig:NGC4388}
% \end{figure}

\section{Data}
\label{sect:data}
 Our data contains 3 cross-dispersed (XD) spectra obtained with: (i) Gemini Near-Infrared Spectrograph \citep[GNIRS,][]{Elias+06a, Elias+06b} on the Gemini North 8.1\,m telescope; (ii) TripleSpec4 \citep{Vacca+03, Cushing+04} mounted on the 4\,m V\'ictor M. Blanco Telescope and (iii) TripleSpec4 on the 4.1-meter Southern Astrophysical Research (SOAR) telescope. We also use integral field unit datacubes obtained with the Spectrograph for INtegral Field Observations in the Near Infrared \citep[SINFONI,][]{Bonnet+03} attached to UT~4, one of the 8\,m telescopes of the Very Large Telescope (VLT), and with the Near-Infrared Integral Field Spectrograph \citep[NIFS,][]{McGregor+03} fed by the Gemini North Adaptive Optics system (ALTAIR). A summary of the observations is made in Tab.~\ref{tab:datasets}.

 For each data set, an A0V star was observed immediately after or before the galaxy, with similar airmass, in order to flux calibrate as well as correct for telluric absorptions. Data reduction was performed using standard reduction scripts for each instrument. These scripts consist of the preparation of the data, flat-field correction, wavelength calibration, telluric atmospheric absorption correction, and flux calibration. For integral field units (IFU), these steps also include spatial calibration and 3D datacube construction.

Since longslit spectra have different slit widths and orientations, we performed two extractions to each IFU, by matching cross-dispersed integrated region. Thus, we created two data sets: one with 0\farcs3$\times$0\farcs3 (covering the years of 2011, 2013 and 2015, hereafter 0\farcs3 data set) and one with 1\farcs1$\times$0\farcs6 (covering the years of 2011, 2015, 2017 and 2021, hereafter 0\farcs6 data set). Also, since NIFS datacube was observed with the adaptive optics, we degraded its spatial resolution to 0\farcs6 before extracting its spectra. Table~\ref{tab:datasets} shows the dates of observation, as well as the instruments, telescopes, and atmospheric conditions.
 
\begin{table} 
    \centering
    \setlength\tabcolsep{4.5pt} % default value: 6pt
    \begin{tabular}{lccccr}
    Date           & instrument & telescope & seeing                    & area                          & Orientation \\
    observed       &            &           & ('') & & ($^{\circ}$E of N)\\\hline
    2011-04-12     & SINFONI   & VLT        &  0.64                     & IFU                           & N/A \\
    2013-02-04     & GNIRS     & Gemini     &  0.49                     & 0\farcs3$\times$0\farcs3      & 64 \\
    2015-07-06     & NIFS      & Gemini     &  0.14 $^1$                & IFU                           & N/A \\
    2017-04-10     & TripleSpec4& Blanco     &  2.5         & 1\farcs1$\times$0\farcs6      & 24 \\
    2021-06-28     & TripleSpec4& SOAR       &  0.80         & 1\farcs1$\times$0\farcs6      & 25
    \end{tabular}
    \caption{Main properties of each data set. $^1$Observed with adaptive optics, and degraded to 0\farcs6.}
    \label{tab:datasets}
\end{table}

Since our data set covers different wavelength ranges, in order to avoid any biases, all analyses conducted throughout this paper are limited between 20000~\r{A} to 24000~\r{A} spectral region, since this is the only wavelength range in common in all data sets. It is worth noting that for XD spectra that extend beyond this region (e.g. covering the bluer spectral region 9000\AA-20000\AA), we also performed the same analysis using the data over the full spectral region, and found no differences beyond our error margin compared to the results presented in this paper. The only significant differences occur in the composition of the stellar population vectors, which we do not present nor discuss.

\begin{figure*}
    \centering
    \includegraphics[width=\textwidth]{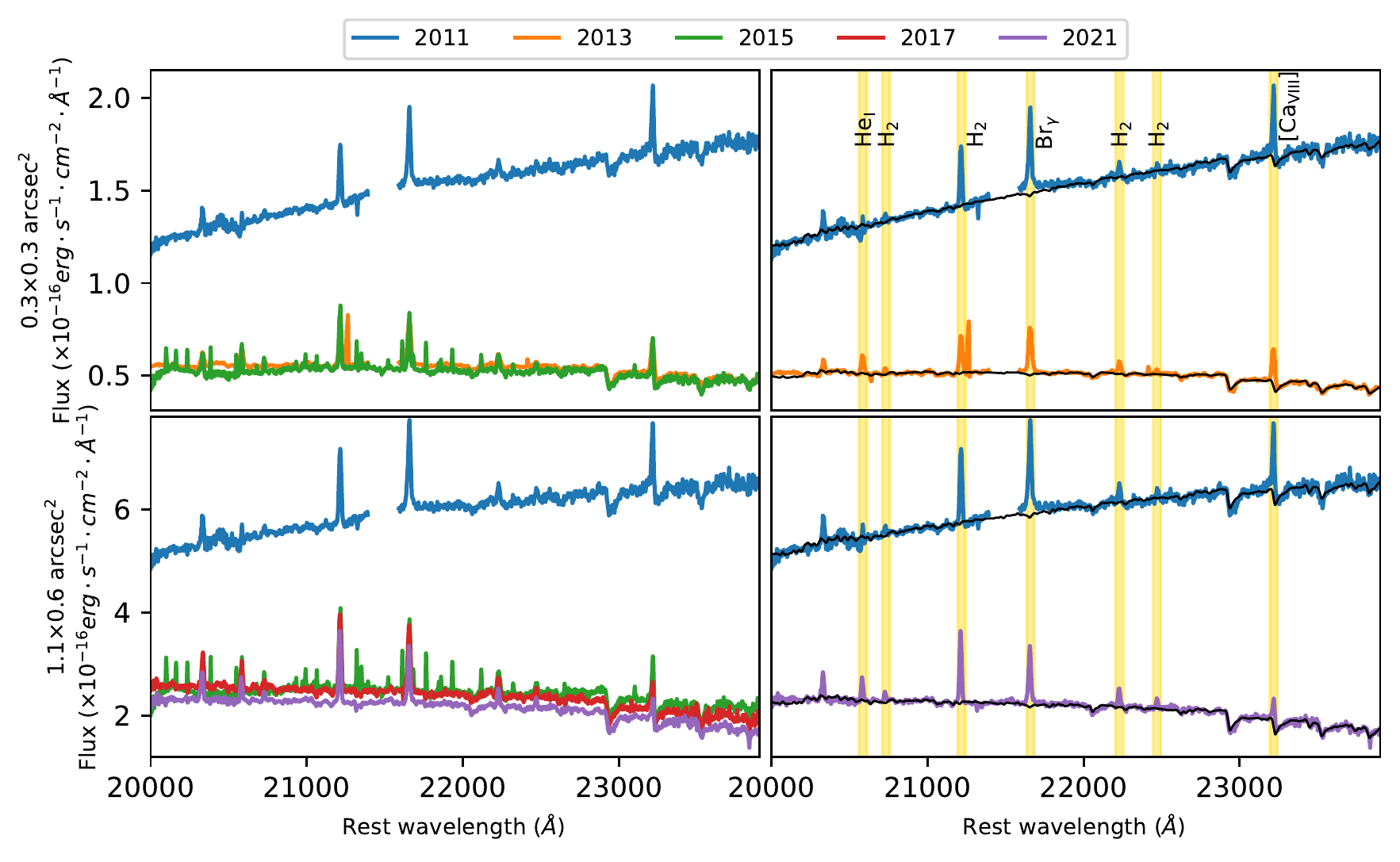}
    \caption{Spectra used in our analysis, where blue, orange, green, red and purple represent those from years 2011, 2013, 2015, 2017 and 2021 respectively. Left panels show all spectra with the same aperture, whereas right shows only two, but with the modelled stellar population in black. At the right panels, we also highlight in yellow the most prominent emission lines. Upper panels show spectra corresponding to an 0\farcs3$\times$0\farcs3 extraction area, whereas bottom panels correspond to extractions of 1\farcs1$\times$0\farcs6.}
    \label{fig:1dspec}
\end{figure*}

Except for the NIFS datacube, all data were observed on non-photometric nights, and thus their flux calibration relative to each other does not match. Because of that, in order to have a better relative flux calibration, we assumed the NIFS datacube as the correct one. For the remaining data sets, we extracted circular regions centred 1\farcs0 away from the nucleus, where the AGN emission is negligible, and then scaled it to their corresponding NIFS flux. This scaling was always performed at $\lambda$22780~\r{A}, since it is far from prominent stellar absorptions as well as emission lines. In the left panel of Fig~\ref{fig:1dspec}, we show the seven spectra used in our analysis. In the upper panels, we show the 0\farcs3 data set, and in the bottom panel we show the 0\farcs6 data set. Also, in the left panels we show two representative spectra and their corresponded spectra as modelled by {\sc starlight}.

Both SINFONI and GNIRS data suffered from poor Br$\gamma$ subtraction in their respective telluric stars, resulting in spurious data in the datacubes around Br$\gamma$. These regions were masked in our analysis. Also, the NIFS datacube suffers from poor subtraction of sky emission lines, which were also masked during our analysis. None of these issues proved to have any significant impact on our analysis, as they were far enough from the emission lines and did not cover much of the continuum region.

\section{Continuum variability}
\label{sect:cont}

In order to disentangle the continuum components we have followed \citet{Riffel+09} and performed stellar population synthesis using the {\sc starlight} code \citep{CF+04,CF+05}. We fed the code with the new generation of the X-shooter Spectral Library of simple stellar population (SSP) models \citep[XSL][]{Verro+22}, created with a \citet{Kroupa01} initial mass function, and PARSEC-COLIBRI isochrones \citep{Bressan+12,Marigo+13}. We limited the library to the nine most representative ages\footnote{50~Myr, 100~Myr, 200~Myr, 500~Myr, 1~Gyr, 2~Gyr, 5~Gyr, 10~Gyr and 15.8~Gyr} and four most representative metallicities\footnote{Z=0.00010, 0.00152, 0.01914 and 0.02409}, totalling 36~SSPs. It is worth mentioning that we have degraded the spectral resolution of the models in order to match that of the data. We also added black body functions (BB) corresponding to temperatures of 700, 800, 1000, 1200 and 1400~K, and a featureless continuum (FC) with f$_\lambda \propto \lambda^{-0.5}$. For a more detailed description of the code and the method of spectral synthesis for NIR data, see \citet{Riffel+09,Riffel+22}. In order to show the quality of the fits, we show in the right panels of Fig~\ref{fig:1dspec} for each data set two spectra with their corresponding synthesis results using {\sc starlight} (black). The most prominent emission lines are highlighted in yellow.

\begin{figure*}
    \centering
    \includegraphics[width=\textwidth]{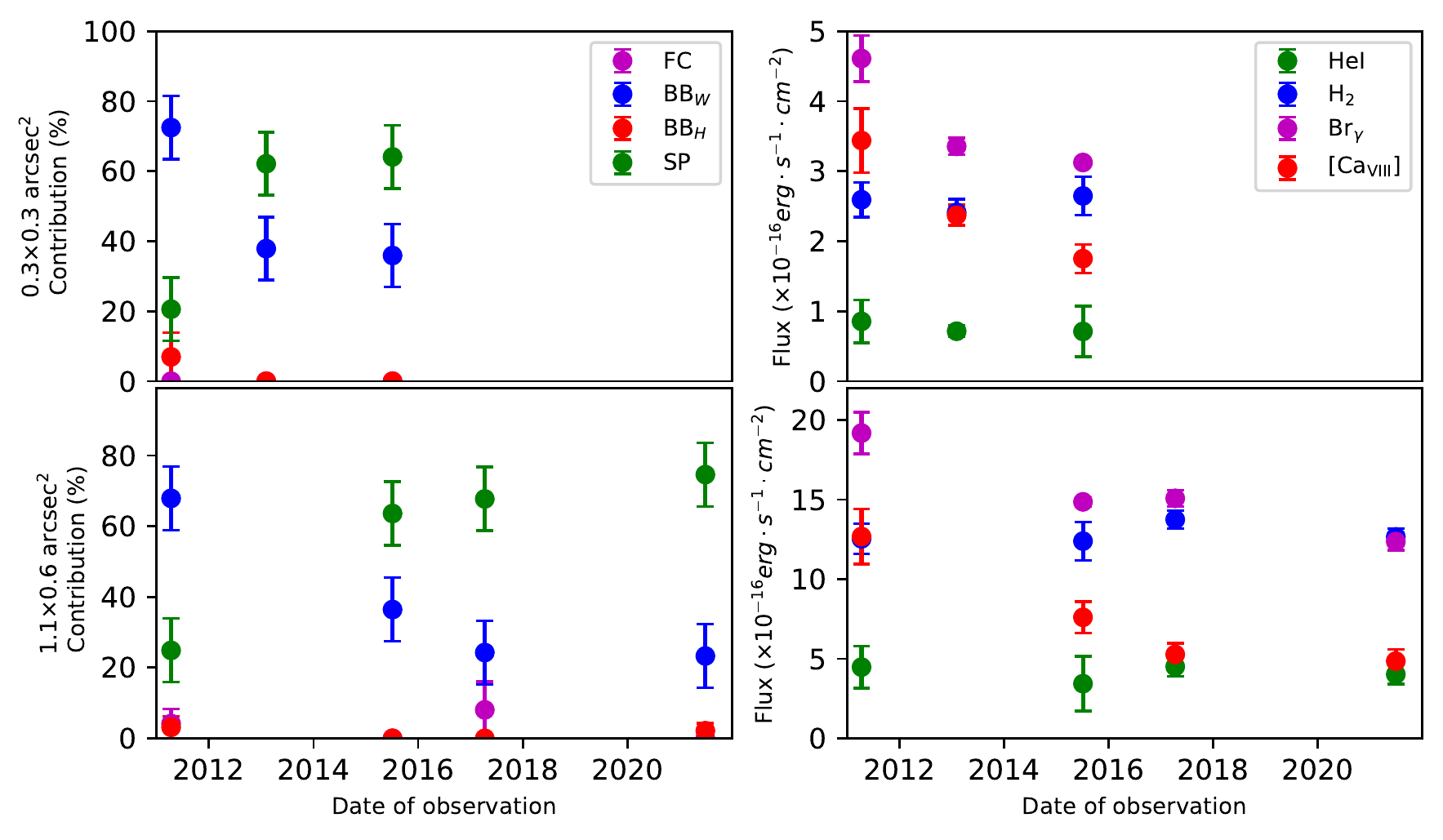}
    \caption{Left: changes in continuum composition between 2011 and 2021 for NGC~4388, as estimated from {\sc starlight}. Right: variations in emission line fluxes in this same period, for the four most  prominent lines. Upper panels show the result for our 0\farcs3 data set, and lower panels show the same results for our 0\farcs6 data set.}
    \label{fig:monitoring}
\end{figure*}

 We divided black bodies into warm (BB$_W$, T$\leq$1000~K) and hot (BB$_H$, T~$>$1000~K). Since our wavelength range is very narrow and excludes J and H bands, where many important stellar absorptions are located, we combined all SSPs in one group, that we have called StPop group. The synthesis results are presented in Table~\ref{tab:synth}, for the percent light contribution at 22780~\r{A}. These same results are presented in Fig~\ref{fig:monitoring}, where we show error bars based on upper limits from \citet{CF+14}. In this paper, the authors estimated {\sc starlight} uncertainties of up to 9~\%, by inducing $\sigma$-level errors on a data set.

\begin{table*}
    \centering
    \begin{tabular}{lccccccc}
 Area &   Date   & FC & BB$_W$  & BB$_H$ & StPop &T$_{BB}$ & BB absolute flux\\   & observed & (\%) & (\%) & (\%)& (\%)& (K) &(erg$\cdot$s$^{-1 }\cdot$cm$^{-2}\cdot$\r{A}$^{-1}$)\\\hline
 \multirow{3}{*}{0\farcs3$\times$0\farcs3}& 
  2011-04-12&  0.   &72.47  &6.96 &20.58 &805$\pm$112&1.19$\times 10 ^{-16}$\\
& 2013-02-04&  0.   &37.87  &0.   &62.12 &825$\pm$118&2.07$\times 10 ^{-17}$\\
& 2015-07-06&  0.   &35.92  &0.   &64.08 &746$\pm$49 &1.90$\times 10 ^{-17}$\\\hline
 \multirow{4}{*}{1\farcs0$\times$0\farcs6}& 
  2011-04-12&  4.17 &67.86  &3.12 &24.86  &822$\pm$124&4.27$\times 10 ^{-16}$\\
&  2015-07-06&  0.   &36.41  &0.   &63.58&756$\pm$49  &8.95$\times 10 ^{-17}$\\
&  2017-04-10&  8.04 &24.26  &0.   &67.7 &873$\pm$110 &5.60$\times 10 ^{-17}$\\
&  2021-06-28&  0.   &23.28  &2.14 &74.58&810$\pm$104 &4.88$\times 10 ^{-17}$\\\hline
    \end{tabular}
    \caption{Continuum composition for spectra taken between 2011 and 2021 for the galaxy NGC~4388. These results correspond to percentage contribution or normalized flux at 22780~\r{A}.}
    \label{tab:synth}
\end{table*}

We can see from Table~\ref{tab:synth} and Fig~\ref{fig:monitoring} that along the monitored period, both hot dust and featureless continuum contributions are negligible in this wavelength range, being lower than our errors in all cases. However, these results also show that between 2011 and 2021 the percent contribution of warm dust decreased by a factor of 2.9 (from 74\% to 25\%) at the 0\farcs6 data set, whereas the contribution of stellar population increased by a factor of 3 (from 68\% to 23\%).

In terms of absolute flux at 22780~\r{A}, the BB$_W$ contribution decreased from 4.27$\times$10$^{-16}$ to 0.49$\times$10$^{-16}$~erg$\cdot$s$^{-1 }\cdot$cm$^{-2}\cdot$\r{A}$^{-1}$ between 2011-04 and 2021-06 in the 0\farcs6 data set, normalized at 22780~\r{A}. This corresponds to an absolute flux decrease of $\sim$88~\%. In the same period, the stellar population flux did not vary beyond our error margin.

\begin{figure*}
    \centering
    \includegraphics[width=\textwidth]{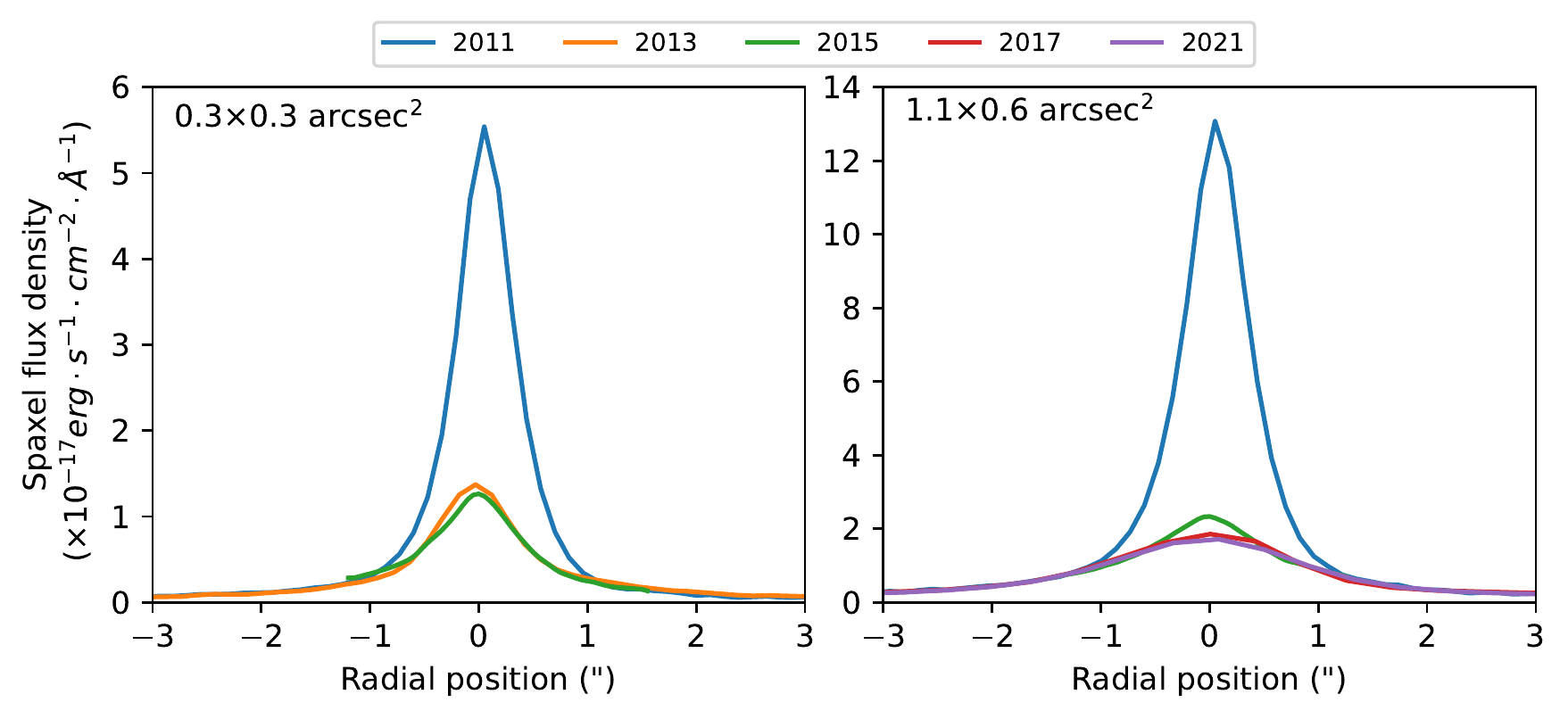}
    \caption{Comparison between the radial profile of the continuum for the five epochs, in both the 0\farcs3 (left) and the 0\farcs6 (right) sample. Since different data have different spaxel sizes, we normalized all data to the NIFS spaxel flux density.}
    \label{fig:profile}
\end{figure*}

In order to better illustrate this change, we measured the radial profile of the continuum at 22780~\r{A}, using a wavelength window of 100~\r{A}, for the two data sets. These results are presented in Fig~\ref{fig:profile}. For SINFONI and NIFS datacubes, we simulated slits with orientations of 64 and 25$^{\circ}$E of N for the 0\farcs3 and 0\farcs6 data sets respectively. Also, since each instrument has different spaxel size, we normalized all radial profiles to the NIFS spaxel flux density.

From Fig~\ref{fig:monitoring}, we can notice that the warm dust contribution fell rather sharply between 2011-04 and 2013-02, decreasing by a factor of ~1.91 in the 0\farcs3 data set. After that initial decrease, it falls within our error margin between 2013-02 and 2015-07 in the 0\farcs3 data set, and then went down by a factor of 1.56 between 2015-07 and 2021-06.  Since our spectra were taken roughly every two years, any changes faster than this threshold would be washed away in our analysis. However, this puts an upper limit of 2~ly (0.6~pc) at the size where this warm dust is located. This scale is compatible with the expected extent of the dusty torus \citep{Zier&Biermann02}.

For each epoch and data set, we estimated the median dust temperature by averaging the BB functions and weighting by luminosity. We found out that the BB temperature remained almost constant within the monitored period, inside our error margin. The average temperature for all observed epochs and data sets is 800$\pm$100~K.  According to \citet{Riffel+09}, we can estimate the upper limit for the mass of the dust (M$_{HD}$) by:

\begin{equation}
    M_{HD} = \frac{4 \pi}{3} a^3 N_{HD} \rho_{gr}
\end{equation}

\noindent where $a$ is the grain radius, $N_{HD}$ is the number of dust grains, and $\rho_{gr}$ is the density of the grain. The number of dust grains can in turn be calculated as $N_{HD}=\frac{L^{HD}_{ir}}{L^{gr}_{\nu, ir}}$, where $L^{HD}_{ir}$ is the total NIR luminosity due to dust, and $L^{gr}_{\nu, ir}$ is the infrared spectral luminosity of each dust grain. 

Also, according to \citet{Riffel+09}, the value of $L^{gr}_{\nu, ir}$ for a typical dust grain at 800~K is 15.08$\times 10^{19}$~erg$\cdot$s$^{-1} \cdot$Hz$^{-1}$, the typical density for graphite grains is 2.26~g$\cdot$cm$^{-3}$, and the typical grain radius is 0.05$\mu$m \citep{Granato&Danese94}.

Thus, by integrating the NIR luminosity due to dust grains, we can see that the total warm dust was 8$\times$10$^{-3}$M$_\odot$ in 2011, but has decreased to 0.9$\times$10$^{-3}$M$_\odot$ in 2021. It means that this AGN had 88\% more warm dust grains in 2011 compared to 2021, but each dust grain emitted roughly the same energy.

\section{Emission line variability}
\label{sect:elines}

We also tracked how the emission lines varied in this same period. We measured the fluxes of the four most prominent emission lines: \ion{He}{I}$\lambda$20587\r{A}, H$_2\lambda$21218\r{A}, \ion{Br}{$\gamma$} and [\ion{Ca}{VIII}]$\lambda$23211\r{A}. We did this by first subtracting the continuum derived in section~\ref{sect:cont}, then measuring the zero flux level by selecting continuum bandpasses to the left and right of the emission line, and finally integrating the area below the emission line.  We notice a false emission feature close to H$_2$ in the GNIRS spectrum obtained in 2013. The spike is very close to the Gaussian profile fit to the molecular line. Thus, when performing the integration, we chose the limits for the emission line and its respective bandpasses aimed at avoiding this flaw. These emission-line fluxes are presented in Table~\ref{tab:elines}, and on the right panel of Fig~\ref{fig:monitoring}.

\begin{table*}
    \centering
    \begin{tabular}{lccccc} 
    \hline
Area & Date observed & \ion{He}{I}& \ion{H}{$_2$} & \ion{Br}{$\gamma$} & [\ion{Ca}{VIII}] \\\hline
 \multirow{3}{*}{0\farcs3$\times$0\farcs3}& 
  2011-04-12& 0.85 $\pm$ 0.30 &  2.59 $\pm$ 0.25 &  4.61 $\pm$ 0.33 &  3.44 $\pm$ 0.45\\
& 2013-02-04& 0.71 $\pm$ 0.07 &  2.41 $\pm$ 0.18 &  3.35 $\pm$ 0.12 &  2.37 $\pm$ 0.14\\
& 2015-07-06& 0.71 $\pm$ 0.36 &  2.65 $\pm$ 0.27 &  3.12 $\pm$ 0.05 &  1.75 $\pm$ 0.20\\\hline
 \multirow{4}{*}{1\farcs1$\times$0\farcs6}& 
  2011-04-12& 4.47  $\pm$ 1.33 & 12.54 $\pm$ 0.94 & 19.18 $\pm$ 1.30 & 12.68 $\pm$ 1.72\\
&  2015-07-06& 3.43 $\pm$ 1.70 & 12.39 $\pm$ 1.20 & 14.86 $\pm$ 0.26 &  7.60 $\pm$ 0.99\\
&  2017-04-10& 4.51 $\pm$ 0.62 & 13.75 $\pm$ 0.56 & 15.08 $\pm$ 0.52 &  5.27 $\pm$ 0.70\\
&  2021-06-28& 4.02 $\pm$ 0.62 & 12.66 $\pm$ 0.52 & 12.36 $\pm$ 0.54 &  4.85 $\pm$ 0.74\\\hline
    \end{tabular}
    \caption{Emission line fluxes of the four most  prominent emission lines in the K band. All fluxes and uncertainties are in units of 10$^{-16}$erg$\cdot$s$^{-1} \cdot$cm$^{-2} \cdot$\r{A}$^{-1}$.}
    \label{tab:elines}
\end{table*}

Within the 10 years monitored period, we detected no variation beyond our error margin in both \ion{He}{I} and H$_2$. On the other hand, the atomic line fluxes of Br$\gamma$ and [\ion{Ca}{VIII}] decreased in the monitored period, 35~\% and  61~\% respectively. 

However, besides these changes being smaller than the 88~\% warm dust variability in this same period, they were also slower. Between 2011-04 and 2013-02, Br$\gamma$ and [\ion{Ca}{VIII}] varied by 27\% and 31~\% in the inner 0\farcs3$\times$0\farcs3, respectively. Normalising that by the total percent variation between 2011 and 2021 (derived from the 0\farcs6 data set), this corresponds to $\sim$77~\% and 49~\% of the total variation in these 10 years. On the other hand, in this same period the warm dust flux decreased by 93~\% of the total variation in the monitored period.% After this initial decrease, between 2013-02 and 2015-07, Br$\gamma$ and [\ion{Ca}{VIII}] varied 7 and 26~\% in our 0\farcs3 data set, with a further decrease of  17 and 36~\% between 2015-07 and 2021-06 in our 0\farcs6 data set respectively.

These results suggest that the warm dust in NGC\,4388 is more internal than the region where Br$\gamma$ and [\ion{Ca}{VIII}] are produced. As can be seen in Tab~\ref{tab:elines}, most of the [\ion{Ca}{VIII}] variation happened between 2011 and 2017, with the flux remaining almost constant after that. Although both ions show spatially resolved emission, which will be discussed in a future paper, our result shows that the bulk of the [\ion{Ca}{VIII}] emission is produced in the inner 2~pc ($\sim$6~ly), as opposed to 0.6~pc for the warm dust. While the warm dust emission is likely produced by the torus itself, either in its inner or its outer walls, [\ion{Ca}{VIII}] is produced in a larger region, possibly in the direction of the ionisation cone.

Br$\gamma$, on the other hand, still presented a variation of 18~\% between 2017-04 and 2021-06, which means it is more extended than [\ion{Ca}{VIII}], spanning a region bigger than 10~ly ($\sim$3~pc). Lastly, the \ion{He}{I} and H$_2$ are even more external, with the bulk of their emission being produced by the host galaxy \citep{RodriguezArdila+05,Riffel+13}, rather than the central source.

There are three differences between our work and the reverberation mapping studies of \citet{Landt+19,Landt+23}: first, they aimed at Seyfert~1 galaxies, whereas we analysed an object where we cannot directly access the central source. Second, their estimated temperature for the dust in the nuclear region lies in the range 1000-1500~K.  Lastly, they detected variations on timescales of months, while we detected variations spanning one decade. Since our spectra were obtained roughly every two years, we do not have enough information to rule out variabilities on shorter timescales, and therefore all our results are upper limits to these timescales and sizes. However, if variations on shorter timescales were present, we would expect to observe more chaotic behaviour between different epochs, such as ups and downs and with varying amplitude. This is in contrast to the smooth variations detected in our analysis, with consistent amplitude and always in one direction. These key differences between our work and the previous reverberation studies in the NIR point to the fact that they were mapping a much more internal region, which is not accessible in NGC~4388. In their study, since the temperatures they derived are close to the one at which dust is expected to sublimate, such dust emission very likely comes from the inner walls of the torus, whereas the emission that we detected is more external than that, coming either from the middle or the outer walls of the torus.

Since this target has confirmed variability in X-rays \citep{Fedorova+11}, they are possibly causing the variations detected in our study. The X-ray variations have timescales of 3-6 months, which means that they are produced in a region closer to the SMBH, no further than 0.15~pc from the central source.  Although these time intervals point to smaller physical scales than the ones probed throughout this paper, according to \citet{Sanfrutos+16} they still track scales associated with the clumpy torus. Given the variability detected in the NIR and its probably association to the X-rays in NGC\,4388, it will be very important to carry out a monitoring study in this object in both spectral regions. It will allow us to set firm constraints to the geometry of both emission regions.

It is worth noticing that we cannot rule out the possibility that these changes are caused by a clump of dust, partially obscuring the warm dust and the emission lines. Nonetheless, the timescales of the variations put a hard constraint on the sizes of the regions detected throughout this paper, since even in the scenario of dust obscuration, the time necessary to eclipse a spectroscopic feature is directly linked to the size of such region.
%\section{Radial profile}
%\label{sect:profile}

% {\color{red} Aqui eu fiz uma figura com as razoes entre os espectros, mas nao gostei muito do resultado. Acho que nao da pra concluir muita coisa, mas deixei aqui para voces verem.}

% \begin{figure}
%     \centering
%     \includegraphics[width=\columnwidth]{ratios.pdf}
%     \caption{Ratios between four spectra and the most recent observation.}
%     \label{fig:ratios}
% \end{figure}

\section{Final remarks}
\label{sect:remarks}

We have monitored the Seyfert~1.9/2 galaxy by means of K-band spectroscopy over a period of 10 years. To the best of our knowledge, it is the first time in the literature that an obscured AGN is monitored in that spectral region. The analysis has allowed us to detect variations in the warm dust emission, never before reported in that source, as well as to constrain the distance of the nuclear dust and the most prominent K-band emission lines in the inner 3~pc. Our main conclusions can be summarised as follows:
\begin{itemize}
    \item The dust emission, which has an average temperature of 800$\pm$100K, decreased 88~\% in flux during the monitoring period. Most of its variation occurred in the first two years, which constrains this emission to the inner 0.6~pc. This puts this emission at the scale of the torus, and since this temperature is much lower than the dust sublimation temperature, this emission likely comes from the middle or outer walls of the torus.
    \item The emission lines of [\ion{Ca}{VIII}] and Br$\gamma$ also varied in this period. Whereas the coronal emission decreased 61~\%, ionised hydrogen decreased 35~\%. However, besides varying less than the warm dust, their variation also took longer. We were able to determine that the bulk of nuclear [\ion{Ca}{VIII}] is produced in the inner 2~pc, with the Br$\gamma$ associated to the central source spanning a region larger than 3~pc.
    \item In the same period, we detected no variation of H$_2$ or \ion{He}{I} beyond our error margin. This suggests that the emission from the AGN does not play a big role in these two lines, with most of their emission probably originating in the host galaxy.
\end{itemize}
Galaxies with variable properties are essential to aid us in constraining the properties of AGN, since the central source cannot be resolved by most telescopes. By adding NGC~4388 to the list of galaxies with confirmed NIR variability, we expect to assist disentangling the geometry of the inner unresolved region of AGNs.

\section*{Acknowledgements}

We thank the anonymous referee for carefully reading the paper and providing suggestions that helped improving the quality of the manuscript. ARA acknowledges financial support from Conselho Nacional de Desenvolvimento Cient\'ifico e Tecnol\'ogico (Proj. ). M.B. acknowledges funding support from program JWST-GO-01717, which was provided by NASA through a grant from the Space Telescope Science Institute, which is operated by the Association of Universities for Research in Astronomy, Inc., under NASA contract NAS 5-03127. RR acknowledges support from the Fundaci\'on Jes\'us Serra and the Instituto de Astrof{\'{i}}sica de Canarias under the Visiting Researcher Programme 2023-2025 agreed between both institutions. RR, also acknowledges support from the ACIISI, Consejer{\'{i}}a de Econom{\'{i}}a, Conocimiento y Empleo del Gobierno de Canarias and the European Regional Development Fund (ERDF) under grant with reference ProID2021010079, and the support through the RAVET project by the grant PID2019-107427GB-C32 from the Spanish Ministry of Science, Innovation and Universities MCIU. This work has also been supported through the IAC project TRACES, which is partially supported through the state budget and the regional budget of the Consejer{\'{i}}a de Econom{\'{i}}a, Industria, Comercio y Conocimiento of the Canary Islands Autonomous Community. RR also thanks to Conselho Nacional de Desenvolvimento Cient\'{i}fico e Tecnol\'ogico  ( CNPq, Proj. 311223/2020-6,  304927/2017-1 and  400352/2016-8), Funda\c{c}\~ao de amparo \`{a} pesquisa do Rio Grande do Sul (FAPERGS, Proj. 16/2551-0000251-7 and 19/1750-2), Coordena\c{c}\~ao de Aperfei\c{c}oamento de Pessoal de N\'{i}vel Superior (CAPES, Proj. 0001). RAR acknowledges financial support from Conselho Nacional de Desenvolvimento Cient\'ifico e Tecnol\'ogico (Proj. 303450/2022-3) and Funda\c c\~ao de Amparo \`a pesquisa do Estado do Rio Grande do Sul (Proj. 21/2551-0002018-0).

%%%%%%%%%%%%%%%%%%%%%%%%%%%%%%%%%%%%%%%%%%%%%%%%%%
\section*{Data Availability}

All data used in this paper are publicly available, and can be freely downloaded through their respective telescope data archives.

%The inclusion of a Data Availability Statement is a requirement for articles published in MNRAS. Data Availability Statements provide a standardised format for readers to understand the availability of data underlying the research results described in the article. The statement may refer to original data generated in the course of the study or to third-party data analysed in the article. The statement should describe and provide means of access, where possible, by linking to the data or providing the required accession numbers for the relevant databases or DOIs.



\bibliographystyle{mnras}
\bibliography{luisgdh} % if your bibtex file is called example.bib

\begin{thebibliography}{}
\makeatletter
\relax
\def\mn@urlcharsother{\let\do\@makeother \do\$\do\&\do\#\do\^\do\_\do\%\do\~}
\def\mn@doi{\begingroup\mn@urlcharsother \@ifnextchar [ {\mn@doi@}
  {\mn@doi@[]}}
\def\mn@doi@[#1]#2{\def\@tempa{#1}\ifx\@tempa\@empty \href
  {http://dx.doi.org/#2} {doi:#2}\else \href {http://dx.doi.org/#2} {#1}\fi
  \endgroup}
\def\mn@eprint#1#2{\mn@eprint@#1:#2::\@nil}
\def\mn@eprint@arXiv#1{\href {http://arxiv.org/abs/#1} {{\tt arXiv:#1}}}
\def\mn@eprint@dblp#1{\href {http://dblp.uni-trier.de/rec/bibtex/#1.xml}
  {dblp:#1}}
\def\mn@eprint@#1:#2:#3:#4\@nil{\def\@tempa {#1}\def\@tempb {#2}\def\@tempc
  {#3}\ifx \@tempc \@empty \let \@tempc \@tempb \let \@tempb \@tempa \fi \ifx
  \@tempb \@empty \def\@tempb {arXiv}\fi \@ifundefined
  {mn@eprint@\@tempb}{\@tempb:\@tempc}{\expandafter \expandafter \csname
  mn@eprint@\@tempb\endcsname \expandafter{\@tempc}}}

\bibitem[\protect\citeauthoryear{{Bischetti} et~al.,}{{Bischetti}
  et~al.}{2017}]{Bischetti+17}
{Bischetti} M.,  et~al., 2017, \mn@doi [\aap] {10.1051/0004-6361/201629301},
  \href {https://ui.adsabs.harvard.edu/abs/2017A&A...598A.122B} {598, A122}

\bibitem[\protect\citeauthoryear{{Bonnet} et~al.,}{{Bonnet}
  et~al.}{2003}]{Bonnet+03}
{Bonnet} H.,  et~al., 2003, in {Wizinowich} P.~L.,  {Bonaccini} D.,  eds,
  Society of Photo-Optical Instrumentation Engineers (SPIE) Conference Series
  Vol. 4839, Adaptive Optical System Technologies II. pp 329--343,
  \mn@doi{10.1117/12.457060}

\bibitem[\protect\citeauthoryear{{Bressan}, {Marigo}, {Girardi}, {Salasnich},
  {Dal Cero}, {Rubele}  \& {Nanni}}{{Bressan} et~al.}{2012}]{Bressan+12}
{Bressan} A.,  {Marigo} P.,  {Girardi} L.,  {Salasnich} B.,  {Dal Cero} C.,
  {Rubele} S.,   {Nanni} A.,  2012, \mn@doi [\mnras]
  {10.1111/j.1365-2966.2012.21948.x}, \href
  {http://adsabs.harvard.edu/abs/2012MNRAS.427..127B} {427, 127}

\bibitem[\protect\citeauthoryear{{Burke} et~al.,}{{Burke}
  et~al.}{2021}]{Burke+21}
{Burke} C.~J.,  et~al., 2021, \mn@doi [Science] {10.1126/science.abg9933},
  \href {https://ui.adsabs.harvard.edu/abs/2021Sci...373..789B} {373, 789}

\bibitem[\protect\citeauthoryear{{Cackett}, {Fabian}, {Zogbhi}, {Kara},
  {Reynolds}  \& {Uttley}}{{Cackett} et~al.}{2013}]{Cackett+13}
{Cackett} E.~M.,  {Fabian} A.~C.,  {Zogbhi} A.,  {Kara} E.,  {Reynolds} C.,
  {Uttley} P.,  2013, \mn@doi [\apjl] {10.1088/2041-8205/764/1/L9}, \href
  {https://ui.adsabs.harvard.edu/abs/2013ApJ...764L...9C} {764, L9}

\bibitem[\protect\citeauthoryear{{Cid Fernandes}, {Gu}, {Melnick}, {Terlevich},
  {Terlevich}, {Kunth}, {Rodrigues Lacerda}  \& {Joguet}}{{Cid Fernandes}
  et~al.}{2004}]{CF+04}
{Cid Fernandes} R.,  {Gu} Q.,  {Melnick} J.,  {Terlevich} E.,  {Terlevich} R.,
  {Kunth} D.,  {Rodrigues Lacerda} R.,   {Joguet} B.,  2004, \mn@doi [\mnras]
  {10.1111/j.1365-2966.2004.08321.x}, \href
  {http://adsabs.harvard.edu/abs/2004MNRAS.355..273C} {355, 273}

\bibitem[\protect\citeauthoryear{{Cid Fernandes}, {Mateus}, {Sodr{\'e}},
  {Stasi{\'n}ska}  \& {Gomes}}{{Cid Fernandes} et~al.}{2005}]{CF+05}
{Cid Fernandes} R.,  {Mateus} A.,  {Sodr{\'e}} L.,  {Stasi{\'n}ska} G.,
  {Gomes} J.~M.,  2005, \mn@doi [\mnras] {10.1111/j.1365-2966.2005.08752.x},
  \href {http://adsabs.harvard.edu/abs/2005MNRAS.358..363C} {358, 363}

\bibitem[\protect\citeauthoryear{{Cid Fernandes} et~al.,}{{Cid Fernandes}
  et~al.}{2014}]{CF+14}
{Cid Fernandes} R.,  et~al., 2014, \mn@doi [\aap]
  {10.1051/0004-6361/201321692}, \href
  {https://ui.adsabs.harvard.edu/abs/2014A&A...561A.130C} {561, A130}

\bibitem[\protect\citeauthoryear{{Cushing}, {Vacca}  \& {Rayner}}{{Cushing}
  et~al.}{2004}]{Cushing+04}
{Cushing} M.~C.,  {Vacca} W.~D.,   {Rayner} J.~T.,  2004, \mn@doi [\pasp]
  {10.1086/382907}, \href {http://adsabs.harvard.edu/abs/2004PASP..116..362C}
  {116, 362}

\bibitem[\protect\citeauthoryear{{Damas-Segovia} et~al.,}{{Damas-Segovia}
  et~al.}{2016}]{Damas-Segovia+16}
{Damas-Segovia} A.,  et~al., 2016, \mn@doi [\apj] {10.3847/0004-637X/824/1/30},
  \href {https://ui.adsabs.harvard.edu/abs/2016ApJ...824...30D} {824, 30}

\bibitem[\protect\citeauthoryear{{Elias}, {Rodgers}, {Joyce}, {Lazo},
  {Doppmann}, {Winge}  \& {Rodr{\'\i}guez-Ardila}}{{Elias}
  et~al.}{2006a}]{Elias+06a}
{Elias} J.~H.,  {Rodgers} B.,  {Joyce} R.~R.,  {Lazo} M.,  {Doppmann} G.,
  {Winge} C.,   {Rodr{\'\i}guez-Ardila} A.,  2006a, in {McLean} I.~S.,  {Iye}
  M.,  eds,  Society of Photo-Optical Instrumentation Engineers (SPIE)
  Conference Series Vol. 6269, Society of Photo-Optical Instrumentation
  Engineers (SPIE) Conference Series. p. 626914, \mn@doi{10.1117/12.671765}

\bibitem[\protect\citeauthoryear{{Elias}, {Joyce}, {Liang}, {Muller}, {Hileman}
   \& {George}}{{Elias} et~al.}{2006b}]{Elias+06b}
{Elias} J.~H.,  {Joyce} R.~R.,  {Liang} M.,  {Muller} G.~P.,  {Hileman} E.~A.,
   {George} J.~R.,  2006b, in {McLean} I.~S.,  {Iye} M.,  eds,  Society of
  Photo-Optical Instrumentation Engineers (SPIE) Conference Series Vol. 6269,
  Society of Photo-Optical Instrumentation Engineers (SPIE) Conference Series.
  p. 62694C, \mn@doi{10.1117/12.671817}

\bibitem[\protect\citeauthoryear{{Emmanoulopoulos}, {Papadakis},
  {Dov{\v{c}}iak}  \& {McHardy}}{{Emmanoulopoulos}
  et~al.}{2014}]{Emmanoulopoulos+14}
{Emmanoulopoulos} D.,  {Papadakis} I.~E.,  {Dov{\v{c}}iak} M.,   {McHardy}
  I.~M.,  2014, \mn@doi [\mnras] {10.1093/mnras/stu249}, \href
  {https://ui.adsabs.harvard.edu/abs/2014MNRAS.439.3931E} {439, 3931}

\bibitem[\protect\citeauthoryear{{Fedorova}, {Beckmann}, {Neronov}  \&
  {Soldi}}{{Fedorova} et~al.}{2011}]{Fedorova+11}
{Fedorova} E.~V.,  {Beckmann} V.,  {Neronov} A.,   {Soldi} S.,  2011, \mn@doi
  [\mnras] {10.1111/j.1365-2966.2011.19335.x}, \href
  {https://ui.adsabs.harvard.edu/abs/2011MNRAS.417.1140F} {417, 1140}

\bibitem[\protect\citeauthoryear{{Forster}, {Leighly}  \& {Kay}}{{Forster}
  et~al.}{1999}]{Forster+99}
{Forster} K.,  {Leighly} K.~M.,   {Kay} L.~E.,  1999, \mn@doi [\apj]
  {10.1086/307761}, \href
  {https://ui.adsabs.harvard.edu/abs/1999ApJ...523..521F} {523, 521}

\bibitem[\protect\citeauthoryear{{Gaidos} et~al.,}{{Gaidos}
  et~al.}{1996}]{Gaidos+96}
{Gaidos} J.~A.,  et~al., 1996, \mn@doi [\nat] {10.1038/383319a0}, \href
  {https://ui.adsabs.harvard.edu/abs/1996Natur.383..319G} {383, 319}

\bibitem[\protect\citeauthoryear{{Gaskell} \& {Klimek}}{{Gaskell} \&
  {Klimek}}{2003}]{Gaskell+03}
{Gaskell} C.~M.,  {Klimek} E.~S.,  2003, \mn@doi [Astronomical and
  Astrophysical Transactions] {10.1080/1055679031000153851}, \href
  {https://ui.adsabs.harvard.edu/abs/2003A&AT...22..661G} {22, 661}

\bibitem[\protect\citeauthoryear{{Granato} \& {Danese}}{{Granato} \&
  {Danese}}{1994}]{Granato&Danese94}
{Granato} G.~L.,  {Danese} L.,  1994, \mn@doi [\mnras]
  {10.1093/mnras/268.1.235}, \href
  {https://ui.adsabs.harvard.edu/abs/1994MNRAS.268..235G} {268, 235}

\bibitem[\protect\citeauthoryear{{Hovatta}, {Tornikoski}, {Lainela}, {Lehto},
  {Valtaoja}, {Torniainen}, {Aller}  \& {Aller}}{{Hovatta}
  et~al.}{2007}]{Hovatta+07}
{Hovatta} T.,  {Tornikoski} M.,  {Lainela} M.,  {Lehto} H.~J.,  {Valtaoja} E.,
  {Torniainen} I.,  {Aller} M.~F.,   {Aller} H.~D.,  2007, \mn@doi [\aap]
  {10.1051/0004-6361:20077529}, \href
  {https://ui.adsabs.harvard.edu/abs/2007A&A...469..899H} {469, 899}

\bibitem[\protect\citeauthoryear{{Hummel} \& {Saikia}}{{Hummel} \&
  {Saikia}}{1991}]{Hummel&Saikia91}
{Hummel} E.,  {Saikia} D.~J.,  1991, \aap, \href
  {https://ui.adsabs.harvard.edu/abs/1991A&A...249...43H} {249, 43}

\bibitem[\protect\citeauthoryear{{Koratkar} \& {Blaes}}{{Koratkar} \&
  {Blaes}}{1999}]{Koratkar&Blaes99}
{Koratkar} A.,  {Blaes} O.,  1999, \mn@doi [\pasp] {10.1086/316294}, \href
  {http://adsabs.harvard.edu/abs/1999PASP..111....1K} {111, 1}

\bibitem[\protect\citeauthoryear{{Koshida} et~al.,}{{Koshida}
  et~al.}{2014}]{Koshida+14}
{Koshida} S.,  et~al., 2014, \mn@doi [\apj] {10.1088/0004-637X/788/2/159},
  \href {https://ui.adsabs.harvard.edu/abs/2014ApJ...788..159K} {788, 159}

\bibitem[\protect\citeauthoryear{{Kroupa}}{{Kroupa}}{2001}]{Kroupa01}
{Kroupa} P.,  2001, \mn@doi [\mnras] {10.1046/j.1365-8711.2001.04022.x}, \href
  {http://adsabs.harvard.edu/abs/2001MNRAS.322..231K} {322, 231}

\bibitem[\protect\citeauthoryear{{Kuo} et~al.,}{{Kuo} et~al.}{2011}]{Kuo+11}
{Kuo} C.~Y.,  et~al., 2011, \mn@doi [\apj] {10.1088/0004-637X/727/1/20}, \href
  {https://ui.adsabs.harvard.edu/abs/2011ApJ...727...20K} {727, 20}

\bibitem[\protect\citeauthoryear{{Landt} et~al.,}{{Landt}
  et~al.}{2019}]{Landt+19}
{Landt} H.,  et~al., 2019, \mn@doi [\mnras] {10.1093/mnras/stz2212}, \href
  {https://ui.adsabs.harvard.edu/abs/2019MNRAS.489.1572L} {489, 1572}

\bibitem[\protect\citeauthoryear{{Landt} et~al.,}{{Landt}
  et~al.}{2023}]{Landt+23}
{Landt} H.,  et~al., 2023, \mn@doi [arXiv e-prints]
  {10.48550/arXiv.2302.01678}, \href
  {https://ui.adsabs.harvard.edu/abs/2023arXiv230201678L} {p. arXiv:2302.01678}

\bibitem[\protect\citeauthoryear{{Marigo}, {Bressan}, {Nanni}, {Girardi}  \&
  {Pumo}}{{Marigo} et~al.}{2013}]{Marigo+13}
{Marigo} P.,  {Bressan} A.,  {Nanni} A.,  {Girardi} L.,   {Pumo} M.~L.,  2013,
  \mn@doi [\mnras] {10.1093/mnras/stt1034}, \href
  {https://ui.adsabs.harvard.edu/abs/2013MNRAS.434..488M} {434, 488}

\bibitem[\protect\citeauthoryear{{McGregor} et~al.,}{{McGregor}
  et~al.}{2003}]{McGregor+03}
{McGregor} P.~J.,  et~al., 2003, in {Iye} M.,  {Moorwood} A.~F.~M.,  eds,
  \procspie Vol. 4841, Instrument Design and Performance for Optical/Infrared
  Ground-based Telescopes. pp 1581--1591, \mn@doi{10.1117/12.459448}

\bibitem[\protect\citeauthoryear{{McHardy}}{{McHardy}}{2001}]{McHardy01}
{McHardy} I.~M.,  2001, in {Peterson} B.~M.,  {Pogge} R.~W.,   {Polidan} R.~S.,
   eds,  Astronomical Society of the Pacific Conference Series Vol. 224,
  Probing the Physics of Active Galactic Nuclei. p.~205

\bibitem[\protect\citeauthoryear{{McHardy}, {Koerding}, {Knigge}, {Uttley}  \&
  {Fender}}{{McHardy} et~al.}{2006}]{McHardy+06}
{McHardy} I.~M.,  {Koerding} E.,  {Knigge} C.,  {Uttley} P.,   {Fender} R.~P.,
  2006, \mn@doi [\nat] {10.1038/nature05389}, \href
  {https://ui.adsabs.harvard.edu/abs/2006Natur.444..730M} {444, 730}

\bibitem[\protect\citeauthoryear{{McHardy} et~al.,}{{McHardy}
  et~al.}{2014}]{McHardy+14}
{McHardy} I.~M.,  et~al., 2014, \mn@doi [\mnras] {10.1093/mnras/stu1636}, \href
  {https://ui.adsabs.harvard.edu/abs/2014MNRAS.444.1469M} {444, 1469}

\bibitem[\protect\citeauthoryear{{Mehdipour} et~al.,}{{Mehdipour}
  et~al.}{2017}]{Mehdipour+17}
{Mehdipour} M.,  et~al., 2017, \mn@doi [\aap] {10.1051/0004-6361/201731175},
  \href {https://ui.adsabs.harvard.edu/abs/2017A&A...607A..28M} {607, A28}

\bibitem[\protect\citeauthoryear{{Netzer}}{{Netzer}}{2015}]{Netzer15}
{Netzer} H.,  2015, \mn@doi [\araa] {10.1146/annurev-astro-082214-122302},
  \href {https://ui.adsabs.harvard.edu/abs/2015ARA&A..53..365N} {53, 365}

\bibitem[\protect\citeauthoryear{{Padovani} et~al.,}{{Padovani}
  et~al.}{2017}]{Padovani+17}
{Padovani} P.,  et~al., 2017, \mn@doi [\aapr] {10.1007/s00159-017-0102-9},
  \href {https://ui.adsabs.harvard.edu/abs/2017A&ARv..25....2P} {25, 2}

\bibitem[\protect\citeauthoryear{{Peterson}}{{Peterson}}{2001}]{Peterson+01}
{Peterson} B.~M.,  2001, in {Aretxaga} I.,  {Kunth} D.,   {M{\'u}jica} R.,
  eds, Advanced Lectures on the Starburst-AGN. p.~3 (\mn@eprint {arXiv}
  {astro-ph/0109495}), \mn@doi{10.1142/9789812811318_0002}

\bibitem[\protect\citeauthoryear{{Riffel}, {Pastoriza},
  {Rodr{\'{\i}}guez-Ardila}  \& {Bonatto}}{{Riffel} et~al.}{2009}]{Riffel+09}
{Riffel} R.,  {Pastoriza} M.~G.,  {Rodr{\'{\i}}guez-Ardila} A.,   {Bonatto} C.,
   2009, \mn@doi [\mnras] {10.1111/j.1365-2966.2009.15448.x}, \href
  {http://adsabs.harvard.edu/abs/2009MNRAS.400..273R} {400, 273}

\bibitem[\protect\citeauthoryear{{Riffel}, {Rodr{\'\i}guez-Ardila}, {Aleman},
  {Brotherton}, {Pastoriza}, {Bonatto}  \& {Dors}}{{Riffel}
  et~al.}{2013}]{Riffel+13}
{Riffel} R.,  {Rodr{\'\i}guez-Ardila} A.,  {Aleman} I.,  {Brotherton} M.~S.,
  {Pastoriza} M.~G.,  {Bonatto} C.,   {Dors} O.~L.,  2013, \mn@doi [\mnras]
  {10.1093/mnras/stt026}, \href
  {https://ui.adsabs.harvard.edu/abs/2013MNRAS.430.2002R} {430, 2002}

\bibitem[\protect\citeauthoryear{{Riffel} et~al.,}{{Riffel}
  et~al.}{2022}]{Riffel+22}
{Riffel} R.,  et~al., 2022, \mn@doi [\mnras] {10.1093/mnras/stac740}, \href
  {https://ui.adsabs.harvard.edu/abs/2022MNRAS.512.3906R} {512, 3906}

\bibitem[\protect\citeauthoryear{{Rodr{\'\i}guez-Ardila}, {Riffel}  \&
  {Pastoriza}}{{Rodr{\'\i}guez-Ardila} et~al.}{2005}]{RodriguezArdila+05}
{Rodr{\'\i}guez-Ardila} A.,  {Riffel} R.,   {Pastoriza} M.~G.,  2005, \mn@doi
  [\mnras] {10.1111/j.1365-2966.2005.09638.x}, \href
  {https://ui.adsabs.harvard.edu/abs/2005MNRAS.364.1041R} {364, 1041}

\bibitem[\protect\citeauthoryear{{Rodr{\'\i}guez-Ardila}
  et~al.,}{{Rodr{\'\i}guez-Ardila} et~al.}{2017}]{RodriguezArdila+17}
{Rodr{\'\i}guez-Ardila} A.,  et~al., 2017, \mn@doi [\mnras]
  {10.1093/mnras/stw2642}, \href
  {https://ui.adsabs.harvard.edu/abs/2017MNRAS.465..906R} {465, 906}

\bibitem[\protect\citeauthoryear{{S{\'a}nchez} et~al.,}{{S{\'a}nchez}
  et~al.}{2017}]{Sanchez+17}
{S{\'a}nchez} P.,  et~al., 2017, \mn@doi [\apj] {10.3847/1538-4357/aa9188},
  \href {https://ui.adsabs.harvard.edu/abs/2017ApJ...849..110S} {849, 110}

\bibitem[\protect\citeauthoryear{{Sanfrutos}, {Miniutti}, {Dov{\v{c}}iak}  \&
  {Ag{\'\i}s-Gonz{\'a}lez}}{{Sanfrutos} et~al.}{2016}]{Sanfrutos+16}
{Sanfrutos} M.,  {Miniutti} G.,  {Dov{\v{c}}iak} M.,   {Ag{\'\i}s-Gonz{\'a}lez}
  B.,  2016, \mn@doi [Astronomische Nachrichten] {10.1002/asna.201612345},
  \href {https://ui.adsabs.harvard.edu/abs/2016AN....337..546S} {337, 546}

\bibitem[\protect\citeauthoryear{{Stone}, {Wilson}  \& {Ward}}{{Stone}
  et~al.}{1988}]{Stone+88}
{Stone} John~L. J.,  {Wilson} A.~S.,   {Ward} M.~J.,  1988, \mn@doi [\apj]
  {10.1086/166458}, \href
  {https://ui.adsabs.harvard.edu/abs/1988ApJ...330..105S} {330, 105}

\bibitem[\protect\citeauthoryear{{Storchi-Bergmann} \&
  {Schnorr-M{\"u}ller}}{{Storchi-Bergmann} \&
  {Schnorr-M{\"u}ller}}{2019}]{StorchiBergmann&SchnorrMuller19}
{Storchi-Bergmann} T.,  {Schnorr-M{\"u}ller} A.,  2019, \mn@doi [Nature
  Astronomy] {10.1038/s41550-018-0611-0}, \href
  {https://ui.adsabs.harvard.edu/abs/2019NatAs...3...48S} {3, 48}

\bibitem[\protect\citeauthoryear{{Vacca}, {Cushing}  \& {Rayner}}{{Vacca}
  et~al.}{2003}]{Vacca+03}
{Vacca} W.~D.,  {Cushing} M.~C.,   {Rayner} J.~T.,  2003, \mn@doi [\pasp]
  {10.1086/346193}, \href
  {https://ui.adsabs.harvard.edu/abs/2003PASP..115..389V} {115, 389}

\bibitem[\protect\citeauthoryear{{Veilleux}, {Bland-Hawthorn}  \&
  {Cecil}}{{Veilleux} et~al.}{1999}]{Veilleux+99}
{Veilleux} S.,  {Bland-Hawthorn} J.,   {Cecil} G.,  1999, \mn@doi [\aj]
  {10.1086/301095}, \href
  {https://ui.adsabs.harvard.edu/abs/1999AJ....118.2108V} {118, 2108}

\bibitem[\protect\citeauthoryear{{Verro} et~al.,}{{Verro}
  et~al.}{2022}]{Verro+22}
{Verro} K.,  et~al., 2022, \mn@doi [\aap] {10.1051/0004-6361/202142387}, \href
  {https://ui.adsabs.harvard.edu/abs/2022A&A...661A..50V} {661, A50}

\bibitem[\protect\citeauthoryear{{Vollmer}, {Pappalardo}, {Soida}  \&
  {Lan{\c{c}}on}}{{Vollmer} et~al.}{2018}]{Vollmer+18}
{Vollmer} B.,  {Pappalardo} C.,  {Soida} M.,   {Lan{\c{c}}on} A.,  2018,
  \mn@doi [\aap] {10.1051/0004-6361/201731910}, \href
  {https://ui.adsabs.harvard.edu/abs/2018A&A...620A.108V} {620, A108}

\bibitem[\protect\citeauthoryear{{Winge}, {Peterson}, {Horne}, {Pogge},
  {Pastoriza}  \& {Storchi-Bergmann}}{{Winge} et~al.}{1995}]{Winge+95}
{Winge} C.,  {Peterson} B.~M.,  {Horne} K.,  {Pogge} R.~W.,  {Pastoriza} M.~G.,
    {Storchi-Bergmann} T.,  1995, \mn@doi [\apj] {10.1086/175730}, \href
  {https://ui.adsabs.harvard.edu/abs/1995ApJ...445..680W} {445, 680}

\bibitem[\protect\citeauthoryear{{Winge}, {Peterson}, {Pastoriza}  \&
  {Storchi-Bergmann}}{{Winge} et~al.}{1996}]{Winge+96}
{Winge} C.,  {Peterson} B.~M.,  {Pastoriza} M.~G.,   {Storchi-Bergmann} T.,
  1996, \mn@doi [\apj] {10.1086/177812}, \href
  {https://ui.adsabs.harvard.edu/abs/1996ApJ...469..648W} {469, 648}

\bibitem[\protect\citeauthoryear{{Yoshida} et~al.,}{{Yoshida}
  et~al.}{2002}]{Yoshida+02}
{Yoshida} M.,  et~al., 2002, \mn@doi [\apj] {10.1086/338353}, \href
  {https://ui.adsabs.harvard.edu/abs/2002ApJ...567..118Y} {567, 118}

\bibitem[\protect\citeauthoryear{{Zier} \& {Biermann}}{{Zier} \&
  {Biermann}}{2002}]{Zier&Biermann02}
{Zier} C.,  {Biermann} P.~L.,  2002, \mn@doi [\aap]
  {10.1051/0004-6361:20021339}, \href
  {http://adsabs.harvard.edu/abs/2002A%26A...396...91Z} {396, 91}

\makeatother
\end{thebibliography}

%\begin{thebibliography}{99}
%\end{thebibliography}

%%%%%%%%%%%%%%%%%%%%%%%%%%%%%%%%%%%%%%%%%%%%%%%%%%

%%%%%%%%%%%%%%%%% APPENDICES %%%%%%%%%%%%%%%%%%%%%

\appendix

%%%%%%%%%%%%%%%%%%%%%%%%%%%%%%%%%%%%%%%%%%%%%%%%%%

% Don't change these lines
\bsp	% typesetting comment
\label{lastpage}
\end{document}